\documentclass{aa}  

\usepackage{graphicx}
\usepackage[normalem]{ulem}
\usepackage{txfonts}
\usepackage{lipsum}
\usepackage{subcaption}         
\usepackage{lscape}             
                                
\usepackage{placeins}

\newcommand{\be}{\begin{equation}}           
\newcommand{\ee}{\end{equation}}

\makeindex

\begin{document}

   \title{Proton acceleration at interplanetary shocks: A numerical study for the HENON mission}

   \author{L. Scarivaglione\inst{1}\fnmsep\thanks{Corresponding author: luisa.scarivaglione@unical.it}
        \and G. Prete\inst{1}
        \and S. Perri\inst{1,2}
        \and G. Zimbardo\inst{1,2} 
        }

   \institute{University of Calabria, Department of Physics, 87036 Rende, Italy
   \and INAF associate, University of Calabria, Rende, Italy}

   \date{}

  \abstract
   {The HENON CubeSat mission of ESA-ASI is designed to fly on a distant retrograde orbit at around 0.08 au upstream of Earth, for advance-time space-weather monitoring. The mission has a set of instruments that include an energetic particle detector measuring protons from 2 MeV to hundreds of MeV. {Detecting energetic particle fluxes well in advance with respect to their impact on Earth is of pivotal importance for protecting technological infrastructures.} 
   }
{In order to assess the features of energetic protons being  accelerated at interplanetary shocks, we aim to set up a test-particle numerical simulation in which protons move in the drift approximation around a parallel shock transition and are accelerated each time they cross the shock
according to a first-order Fermi process. {During their motion they can propagate around the shock either diffusively or superdiffusively; particle transport indeed affects shock acceleration.}} 
   {Protons were injected at the shock with an energy distribution
of about 10 keV and scattered in pitch angle by a collision operator. In the simulations, we varied the
pitch-angle scattering time, the plasma speeds upstream and downstream of the shock, and the type
of transport. In the superdiffusive case, a
power-law distribution of scattering times was generated by mimicking a L\'evy walk. Several runs were carried out in order to determine the parameter domain, which led to
efficient acceleration {and thus higher} energies.}
   {For typical parameters,
superdiffusive acceleration turns out to be somewhat faster than normal diffusive acceleration. We find that the
main control parameter, which determines the maximum energy, appears to be the upstream shock speed, and
that energies in the range of 2--10 MeV can be reached for upstream speeds in the shock frame, $V_1,$ of the order of 300 km/s or more. Conversely, for a speed of $V_1 \leq 150$ km/s the proton energy reaches only a few hundred keV. }
   {These results facilitate our understanding of why many interplanetary shocks only accelerate ions up to moderate energies. Furthermore, the present study
indicates that when the HENON particle instrument is able to detect energetic protons above 2 MeV, the shock speed in the spacecraft frame is relatively high, i.e. of the order of 800 km/s or more.}

\titlerunning{Proton acceleration at interplanetary shocks} 

\maketitle
\nolinenumbers

\section{Introduction}

Energetic particles of solar origin with energies up to several MeV and even a few GeV are frequently observed in the heliosphere, and collisionless shock waves are considered to be one of the prime acceleration mechanisms \citep[e.g.][]{Reames99,Lee12}. Besides solar energetic particles (SEPs), which are accelerated in the vicinity of the Sun, the so-called energetic storm particles (ESPs) are accelerated by travelling interplanetary (IP) shocks and are detected when the shock passes over a spacecraft in the solar wind  \citep[e.g.][]{Decker81,Simnett98,Lario03,Giacalone12,Dresing16,Lario18}.
However, not all the shock waves observed in the heliosphere are associated with local (to the shock) particle acceleration; while ESPs are frequently observed for the fastest and strongest shocks \citep[e.g.][]{Giacalone12,Lario18}, only a minority of IP shocks are associated with local acceleration. For instance, using STEREO data \citet{Dresing16} found that there is a shock-associated increase of 100 keV ions for 27\% of shocks, but the increase of 2 MeV ions is only for 5\% of shock crossings. 
As reported in a recent review by \citet{Lugaz24b}, energetic storm particles only occur in 15--35\% of IP shocks and have typical energies from about 100 keV to 5 MeV. Furthermore, there is no simple relationship between the properties (e.g. intensities, maximum energy, spectral index, and time profiles) of ESPs and the shock parameters such as Mach numbers, compression ratios, shock normal angles, and so on \citep{vanNes84,Neugebauer05,Lee12,Trotta25b}. In addition, many IP shocks are driven by coronal mass ejections (CMEs) propagating from the Sun outwards, and,  
as pointed out by \citet{Lee12}, the broad range of SEP 
intensities observed with similar CME speeds is one well-advertised puzzle. 
Given the ubiquity of shocks in the heliosphere and in most astrophysical environments, understanding under what conditions ESPs can be efficiently accelerated at shock waves is now urgent. 

Coronal-mass-ejection-driven shocks and CMEs themselves represent some of the major hazards of space weather. In this connection, a new space mission called HEliospheric pioNeer for sOlar and interplanetary threats defeNce (HENON) is being developed by ESA-ASI under the GSTP programme, with the aim of sending a spacecraft on a distant retrograde orbit (DRO) around the Earth to provide observations upstream of the L1 Lagrangian point \citep{Provinciali24,Cicalo25}.
A sub-L1 monitor allows for in situ observations of space-weather drivers such as shocks and CMEs, with a lead time of several hours compared to those obtained at L1 \citep{Lugaz24a,Lugaz25,Palmerio25}. HENON is a cubesat that will fly on a DRO at 0.082 au upstream of the Earth, 
{which corresponds to about 0.92 au from the Sun. HENON} will take plasma, energetic particles, and magnetic-field measurements to compare with those {obtained by a spacecraft} at L1. The lead time varies between two and eight hours, depending on the solar-wind speed and HENON position along the DRO \citep{Prete26a}. In particular, HENON will carry an energetic particle detector called the Relativistic Electron and Proton Experiment (REPE), which will measure protons from 2 MeV to hundreds of MeV, and electrons from 100 keV to 10.4 MeV \citep{Ngom26}. 
Given that not all IP shocks accelerate particles and not all such ESPs reach the energy threshold of 2 MeV for protons of the REPE instrument, it is necessary to assess which IP shock parameters lead to the acceleration of $>2$ MeV protons. In other words, when we know which shock properties make it possible to reach and overcome 2 MeV   we may try to constrain the shock-parameter range when REPE observes protons associated with an
IP shock. For example, the observation of protons with energy levels of $E>2$ MeV can give us information on the shock speed in the spacecraft frame, the gas compression ratio, and the Mach numbers in the shock frame. 

In this work, we carried out a number of test-particle simulations to explore proton acceleration by a shock transition. Particles were injected at the shock front and scattered back and forth by a collision operator that mimics pitch-angle diffusion; i.e. when protons cross the shock boundary their momentum is increased in accordance with the theory of first-order Fermi acceleration, as assumed in diffusive shock acceleration (DSA; \citealt{bell1978acceleration,Drury83,Gaisser90}). In addition, we explored not only the case of normal diffusion, as assumed in standard DSA, but also the case of superdiffusion, i.e. the cases when the mean square displacement of particles scales with time as
\be
\langle (\Delta x)^2 \rangle = 2 {\cal D}_{\alpha} t^{\alpha}
\label{eq:meanSquareDispl}
,\ee
with $\alpha>1$. Indeed, \citet{Perri07,Perri08}, \citet{Perri15b}, and \citet{prete2021energetic} have shown that for many IP shocks the energetic particle intensity profile is better described by a power-law decay in the upstream region, which is a signature of superdiffusion. Conversely, in the case of normal diffusion, the upstream intensity profile is described by an exponential decay. Both kinds of upstream profile --- i.e. exponential and power-law profiles --- are observed in space, so both transport regimes need to be studied. In the superdiffusive case, a power-law distribution of scattering times needs to be generated, as we show in Sect. 2. In the simulations, we varied parameters such as the upstream and downstream plasma velocities in the shock frame (which also implies a variation of the compression ratio), the kind of transport, i.e. either normal or superdiffusive, and the scattering times (which correspond to varying the diffusion coefficients). In principle, pitch-angle scattering depends on the properties of magnetic turbulence in resonance with the particle gyroradius \citep[e.g.][]{Giacalone99}. Such turbulence properties vary widely \citep[e.g.][]{Pitna21,Perri23,ZhaoS25}, so several different values of the scattering time can be obtained.

We find that shocks with low upstream speed are less efficient in accelerating particles, while energies in the 2--10 MeV range can be reached in a number of cases. We discuss the influence of the scattering times and the shock parameters that lead to efficient proton acceleration. We also discuss the conditions for the detectability threshold by the REPE instrument and consider how the observations of particles with $E>2$ MeV can translate into a pre-alert (to be issued by HENON) of an approaching strong perturbation.

\section{Numerical simulation and choice of model parameters}

To study the dynamics of protons around a planar shock wave, we adopted the test particle numerical code developed by \citet{Prete2019,prete2021energetic} and extended it to the case of acceleration by \citet{Prete26b}. In this one-dimensional code, particles move in the drift approximation along the magnetic field, which is taken to be perpendicular to the planar shock surface. That is, in this work we considered parallel shocks in which the shock normal angle $\theta_{Bn}=0$. This is the same configuration originally considered by \citet{bell1978acceleration}. Particles move in a one-dimensional simulation box in the shock frame, with the shock at rest in $x=0$, and are scattered in pitch-angle, $\theta,$  by a collision operator such that 
\be
v_{\parallel} = v \mu
;\ee
here, $v_{\parallel}$ is the velocity along the magnetic field, $\mu=\cos \theta$ is the pitch-angle cosine, and $v$ is the particle speed, which depends on energy. At each scattering time, $\tau$, a new $\mu$ is obtained as $\mu = 2\xi -1,$ with $\xi$ being a random number evenly distributed between zero and one. This scattering scheme generates isotropic particle distributions. Protons are injected at $x=0$ and advanced in time according to 
\begin{equation}
    dx_i = (V_{bulk} + v_{\parallel})dt,
    \label{eq:langevin}
\end{equation}
where the time step $dt=1$s for most runs; this time step corresponds to the inverse of the proton gyrofrequency, $\Omega_p$, in a typical solar-wind magnetic field of 10 nT at 1 au. Since we did not resolve the gyromotion, it was sufficient to fix $dt\simeq \Omega_p^{-1}$. 
The proton velocity, $v\mu,$ is changed randomly when the current time $t$ increases by an amount corresponding to the scattering time, $\tau$. Instead, the bulk velocity in the shock frame equals $V_1$ in the upstream side of the shock and $V_2$ in the downstream side. The plasma velocities $V_1$ and $V_2$ were varied in the different runs in the way specified below. {The compression ratio, $r$, in the numerical model is not an independent input parameter, it was determined by the upstream and downstream velocities defined in the initial setup of each run. Consequently, our analysis not only investigated the effect of different compression ratios, it also compared shock configurations with the same $r$ and different values of $V_1$ and $V_2$ so as to reveal the role of the individual flow velocities.} 

When a particle crosses the shock, {its average change in momentum, $p$ (averaged over an isotropic pitch-angle distribution),} is given by  
\begin{equation}
\langle \Delta p\rangle=\frac{2}{3} p (V_1-V_2)/v,
\label{eq:momentum}
\end{equation}
which is in agreement with the average momentum gain in first-order Fermi acceleration \citep{Drury83}. {Notice that $p$ and $v$ were evaluated in the upstream fluid frame.} At each shock crossing, the particles’ velocities were first evaluated in the upstream and downstream reference frames and then transformed into the shock reference frame via a Galilean transformation \citep[see][for more details]{Prete26b}. The simulation box for the computation of particle densities and energy spectra extends from $\pm L$, with $L=10^7$ km for comparison with spacecraft observations in the heliosphere \citep{Prete2019}, but particles are allowed to move in a larger region of space to avoid the effects of absorbing boundaries.

The scattering times, $\tau,$ were generated differently in the cases of diffusion and superdiffusion. For normal diffusion, corresponding to standard DSA, we made the simple choice $\tau= \rm const$, with the values of $\tau$ chosen as explained below. Normal diffusion gives rise to an exponential intensity profile upstream of the shock and to a constant profile downstream \citep[e.g.][]{Drury83,Giacalone12}; {this can be obtained as the solution of a stationary advection-diffusion equation, assuming a diffusion coefficient that depends on the particle energy only.} 
When superdiffusion is implemented in the model, i.e. when we model superdiffusive shock acceleration (SSA; \citealt{Perri12a,Zimbardo13}), the particle motion is described by a L\'evy random walk \citep{Klafter87,Zumofen93,Zaburdaev15}. This involves a probability $\Psi$ of free path lengths $\ell$ with a power-law shape \citep{Zimbardo13,Trotta15, Prete2019,Zim20,Aerdker25} of 
\begin{equation}
\Psi (\ell, \tau) =
\left\lbrace
\begin{array}{ll}
\frac{1}{2}C\delta(|\ell| - v\tau), \;\;\;\;\;\;\;  |\ell| < \ell_0 \\
\\
\frac{1}{2}C |{\ell}/{\ell_0}|^{\alpha-4}\delta(|\ell| - v\tau), \;\;\;\;\;\;\;  |\ell| > \ell_0 \, .
\end{array}
\right.
\label{eq:levyWalk}
\end{equation}
Here, $\ell_0$ indicates a scale parameter, namely a characteristic free path length, such that for $\ell>\ell_0$ the probability distribution of the path lengths is a power-law decay with exponent $4-\alpha$; here, $1<\alpha <2$ \citep{Klafter87,Zumofen93,Zaburdaev15}. Notice the presence of the Dirac delta in the expression of $\Psi(\ell, \tau)$, which ensures the coupling between free path lengths and scattering times, so that $|\ell| = v \tau$ and $\ell_0 = v \tau_0$, with $\tau_0$ being a scale time. The exponent $\alpha-4$ quantifies the importance of the power-law tails of $\Psi(\ell, \tau)$, with larger $\alpha$ implying  heavier tails. For superdiffusion, the mean square displacement of particles assumes the form given by Eq. (\ref{eq:meanSquareDispl}) 
for $t \to \infty$, with the exponent $1<\alpha <2$; on the other hand, normal diffusion is restored for $\alpha\to 1$. 
We can generate a sequence of scattering times corresponding to Eq. (\ref{eq:levyWalk}) for $\xi>\xi_0$ as \citep{Trotta15}
\begin{equation}
    \tau = \tau_0 \left[\frac{1}{ (4-\alpha) \left(1 - \xi \right)}\right]^{1/(3 - \alpha)} ,
    \label{eq:scattSSA}
\end{equation}
while $\tau=\xi/C$ for $\xi < \xi_0$ (for more details, see \citealt{Prete2019,prete2021energetic} and \citealt{Prete26b}). When $\xi \to 1 $, very long $\tau$ are allowed, and due to the coupling between space and time, Eq. (\ref{eq:scattSSA}) leads to the emergence of the very long displacements that characterize L\'evy walks.

We notice that superdiffusion gives rise to power-law intensity profiles for energetic particles upstream of shocks, and to an intensity decrease downstream: such types of profiles are frequently observed in space \citep[e.g.][]{Perri08,Perri15b,prete2021energetic}. Furthermore, a recent analysis of ions accelerated at IP shocks observed by Solar Orbiter  \citep{Trotta25b} shows that energetic-particle `spike' profiles (likely due to superdiffusion) are at least as common as plateau profiles (corresponding to normal diffusion) for ions in the energy ranges from 50 keV to 1 MeV. 

The efficiency of shock acceleration depends on several conditions and parameters such as the presence or absence of a population of suprathermal seed particles; the shock speed and the gas compression ratio; the angle, $\theta_{Bn}$, between the shock normal and the upstream magnetic field; the Alfv\'enic and magnetosonic Mach numbers; the interaction with turbulence that scatters protons at a pitch angle; and the total time available for shock acceleration, which in our case corresponds to the shock lifetime. In the present work, we chose the above features in the following way. As mentioned above, the angle $\theta_{Bn}=0$ since we assumed a parallel shock.
The injection energy of protons was modelled according to the properties  of so-called diffuse ions observed in the vicinity of the Earth's bow shock \citep[e.g.][]{Paschmann81,Trattner23}. In the quasi-parallel region of the Earth's bow shock, diffuse ions have energies of tens of keV, with a peak around 10--20 keV \citep{Paschmann81}, an energy tail up to 30-40 keV \citep{Mazelle03}, and a nearly isotropic exponential distribution \citep{Trattner23}. Such suprathermal ions are created by ion reflection at shocks \citep[e.g.][]{Edmiston84}, and are accelerated by wave-particle interaction and shock drift acceleration \citep[e.g.][]{Decker86}. 
In addition, we note that particle-in-cell numerical (PIC) simulations of shock waves are able to accelerate suprathermal particles extracting them directly from the thermal populations. For instance, \citet{Trotta20} found, by running PIC simulations of a quasi-parallel shock with $\theta_{Bn}=15^\circ$, plasma $\beta = 0.5$, and Alfv\'enic Mach number $M_A=8$, that suprathermal non-Maxwellian tails are formed upstream and downstream (see their Fig. 2); those tails have velocities extending up to 80 Alfv\'en speeds. Using parameters typical of the solar wind, this corresponds to energies of up to 40--70 keV. Here, on the basis of the above observations and self-consistent studies, we assumed an injection energy distribution for protons  given by
\be
E = E_0 - \Delta E \ln{\eta} \ ,
\ee
with $\eta$ being a random number evenly distributed in $0<\eta<1$. 
\begin{figure}[ht!]
  \centering
  \includegraphics[width=0.8\linewidth]{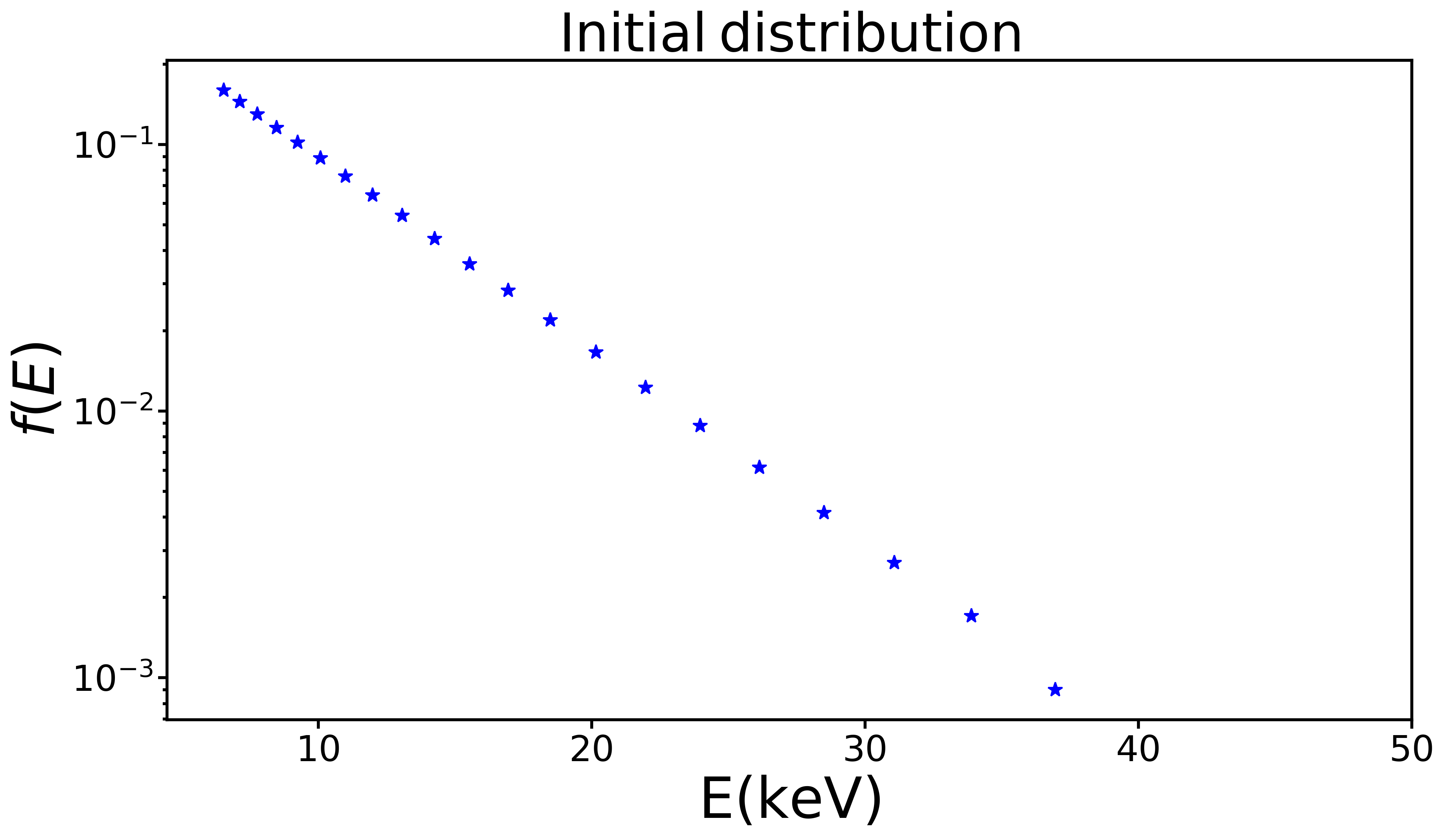}
  \caption{Initial injection energy distribution.}
  \label{init_E0}
\end{figure}
This generates an exponential energy distribution of $f(E)\propto \exp{[-(E-E_0)/\Delta E}]$; for the present runs, we fixed $E_0= 6.25$ keV and $\Delta E=6$ keV, and the resulting initial distribution is shown in Fig. \ref{init_E0}. Limiting the energy peak to 6.25 keV (rather than 10--20 keV as discussed above) is considered appropriate for IP shocks, as they normally have Mach numbers smaller than the bow shock.

The shock speed and compression ratio were varied in our simulation with the aim of identifying the values that can lead to proton acceleration above 2 MeV. Based on the analysis of a number of IP shock crossings, we chose the upstream plasma velocities in the shock frame as $V_1=100$, 150, 200, 250, 300, 350, 400, 450, 500, 600, 700, and 800 km s$^{-1}$;
and the downstream velocities as $V_2= 70$, 100, 150, and 200 km s$^{-1}$. These velocities also determine the compression ratio. For reference, we assumed an upstream plasma speed in the spacecraft frame of $V_1^{\rm s/c}=450$ km s$^{-1}$ (the typical solar-wind speed), while the other velocities were obtained by a Galilean transformation. We worked in the normal incidence frame, ${\bf V}_1 \cdot \hat{x} \equiv V_1$, with $\hat{x}$ being the shock normal.  
Considering that all vectors are in the $x$ direction in our 1D model, we have
\be
V_1^{\rm s/c} = V_{\rm sh}^{\rm s/c} + V_1 \ ,
\ee
\be
V_2^{\rm s/c} = V_{\rm sh}^{\rm s/c} + V_2 \ ,
\ee
where $V_{\rm sh}^{\rm s/c}$ is the shock speed in the spacecraft frame. This was not measured directly but determined by \citep[e.g.][page 162]{Burgess95}
\be 
V_{\rm sh}^{\rm s/c} = \frac{r{\bf V_2^{\rm s/c}}  - {\bf V_1^{\rm s/c}}} {r-1} \cdot \hat{x}
,\ee 
where $r=V_1/V_2=n_2/n_1$ is the compression ratio, with $n$ being the number density, which is usually measured by spacecraft. From these transformations we were able to obtain both the shock speed, $V_{\rm sh}^{\rm s/c}$, and the downstream speed, ${V_2}$. We note that in the shock frame $V_1$ and $V_2$ are actually negative (looking from a forward shock, the plasma is moving `towards the Sun'), but in the discussion and in the figures we consider $V_1$ and $V_2$ as their absolute values. 

The interaction of protons with magnetic turbulence causes pitch-angle scattering and diffusion in the parallel velocity, $v_{\parallel}$, and such a diffusion is one of the main ingredients of shock acceleration. In the case of normal transport, the diffusion coefficient can be written as
\be 
D= \frac{1}{3} \lambda v = \frac{1}{3} v^2 \tau
\label{eqscatt_normal}
,\ee 
where $\lambda=v\tau$ is the mean free path. Since the particle speed $v$ is determined by the energy $E$ (it being either the injection energy or the energy growing because of acceleration), what remains to be determined is the scattering time, $\tau$. Estimates based on the turbulence level in resonance with particles vary widely \citep{Giacalone99,Shalchi07,Perrone13}. Here, we deduced the {scattering times directly from the observations of the upstream flux of energetic particles at a number of IP shocks}. Indeed, the diffusion coefficient along the magnetic field, 
 $D_{\parallel}$, can be obtained from the component of the diffusion tensor along the radial direction (along the normal to the shock), namely $D_{\parallel} = D_{xx}/\cos^2 \theta_{Bn}$. Therefore, one can infer the scattering time from the analysis of the upstream exponential rise of energetic particle fluxes, $J\propto \exp{[V_1 x/D_{xx}]}$ for $x<0$. In particular,  \citet{Giacalone12} found radial diffusion coefficients ranging from $0.6\times 10^{18}$ cm$^2$ s$^{-1}$ to $3.1\times 10^{18}$ cm$^2$ s$^{-1}$ for energetic protons in the energy channels from 47 keV to 310 keV. Considering the corresponding shock normal angles, $\theta_{Bn}$,  and the average velocities in each energy channel, scattering times ranging from 35 s to 150 s can be inferred. Another example is given by \citet{Ding24}, whose authors analysed the September 6, 2022 shock crossing measured by the Solar Orbiter and obtained radial-diffusion coefficients (staying very close to the shock) ranging from $5\times 10^{18}$ cm$^2$ s$^{-1}$ to $1.8\times 10^{19}$ cm$^2$ s$^{-1}$ for energetic protons in the energy channels from 300 keV to 1 MeV. These coefficients led to scattering times from about 60 s to 75 s. Recently, for energetic protons accelerated at shocks observed by the Solar Orbiter,  \citet{Kartavykh25} found mean free paths of the order of 0.013--0.015 au for particles from 130 keV to 1 MeV. The corresponding scattering times, $\tau=\lambda/v,$ are of the order of 140 s. Hence, for the runs modelling normal diffusion, we fixed $\tau$ as 50 s, 100 s, and 150 s, considering these as typical values when particles are indeed accelerated. For simplicity,  
 we kept the same value for the scattering times both upstream and downstream and independent of energy. 

In the case of superdiffusion, \citet{prete2021energetic} found, by analysing Advanced Composition Explorer shock crossings, that the scale time, $\tau_0,$ which separates the power-law part of $\Psi (\ell, \tau)$ from the flat part (see Eq. \ref{eq:levyWalk}) ranges from about 1 s to 10 s (the methodology is outlined in \citet{Perri15b} and in Prete et al. 2026). Here, we fixed $\tau_0=5$ s and 10 s. 
It should be borne in mind that superdiffusion encompasses both long (much longer than $\tau_0$) and short scattering times. 

We can easily grasp the influence of parameters such as the plasma velocities and the scattering times on particle acceleration by recalling the expression of the cycle time for particles moving around a shock from upstream to downstream and back \citep{Lagage83,Lagage83b,Drury83}:
\be
t_{\rm cycle} \simeq \frac{4}{v} \left (\frac{D_1}{V_1} + \frac{D_2}{V_2} \right) \ ,
\label{time_cycle}
\ee 
where $D_1$ and $D_2$ are the upstream and downstream diffusion coefficients, respectively. As shown by Eq. \ref{eqscatt_normal}, for a given particle speed, the shorter the scattering time, $\tau,$ and the larger the bulk velocities, the faster the acceleration process, i.e. $t_{\rm cycle} \sim v\tau/V_1$.
An analogous expression can be obtained for superdiffusive shock acceleration as \citep{Perri12a,Zimbardo13}
\be
T_{\rm cycle} \simeq \frac{4}{v} \biggl[\biggl(\frac{{\cal D}_{\alpha 1}}{V_1^{\alpha}}\biggr)^{1/(2-\alpha)}+\biggl(\frac{{\cal D}_{\alpha 2}}{V_2^{\alpha}}\biggr)^{1/(2-\alpha)}\biggr] \, .
\label{time_cycle_super}
\ee
It can be shown that the anomalous diffusion coefficient,  ${\cal D}_{\alpha}$, scales as ${\cal D_{\alpha}} \propto v^2 \tau_0^{2-\alpha}$ \citep{Perri12a,Zimbardo13,Perri15}, so in the case of superdiffusion the cycle time is proportional to $\tau_0$, i.e. $T_{\rm cycle} \sim (v/V_1)^{\alpha/(2-\alpha)} \, \tau_0$. Therefore, the smaller $\tau_0$, the faster the acceleration. Yet, in the case of superdiffusion, another parameter is relevant, i.e. the exponent of superdiffusion, $\alpha$; we can see that the larger $\alpha$ is, the longer the cycle time is, keeping in mind that the particle velocity, $v,$ is much larger than the bulk velocity, $V_1$. Anyway, the larger $V_1$ is, the shorter the cycle time is, as in the case of normal diffusion but with a larger exponent on the velocity ratio. 

Finally, we had to fix a maximum time for the acceleration process, which in practice corresponds to the shock lifetime if we assume that the shock properties do not change during its propagation. The maximum available time determines the maximum reachable energy, which is a typical property of Fermi acceleration. Indeed, for supernova remnant shocks the maximum cosmic-ray energy (for protons) is limited by the supernova age \citep{Lagage83, Lagage83b}. Here, we consider that fast shocks, which are more likely to be efficient particle accelerators, move at about $800$ km/s (or more), so the travel time from the Sun to the Earth is about two days (i.e. about $1.7\times 10^5$ seconds). In the simulations, we also kept track of the energy spectrum at intermediate times to better quantify the acceleration process. 

\section{Simulation results}
We carried out a campaign of numerical simulations varying the upstream and downstream shock velocities, as well as the scattering times. This parametric study explored both diffusive and superdiffusive shock-crossing regimes for test particles. For all runs, we have the same injection distribution shown in Fig. \ref{init_E0}. By injecting $10^7$ particles in each run, the proton energy spectra are obtained as a function of the elapsed time, and we find that a power-law energy spectrum is formed within the times corresponding to the shock lifetime.  
The full lists of simulation parameters investigated in this study are reported in Appendix A. 

\begin{figure*}[ht!]
  \centering
  \includegraphics[width=0.95\linewidth]{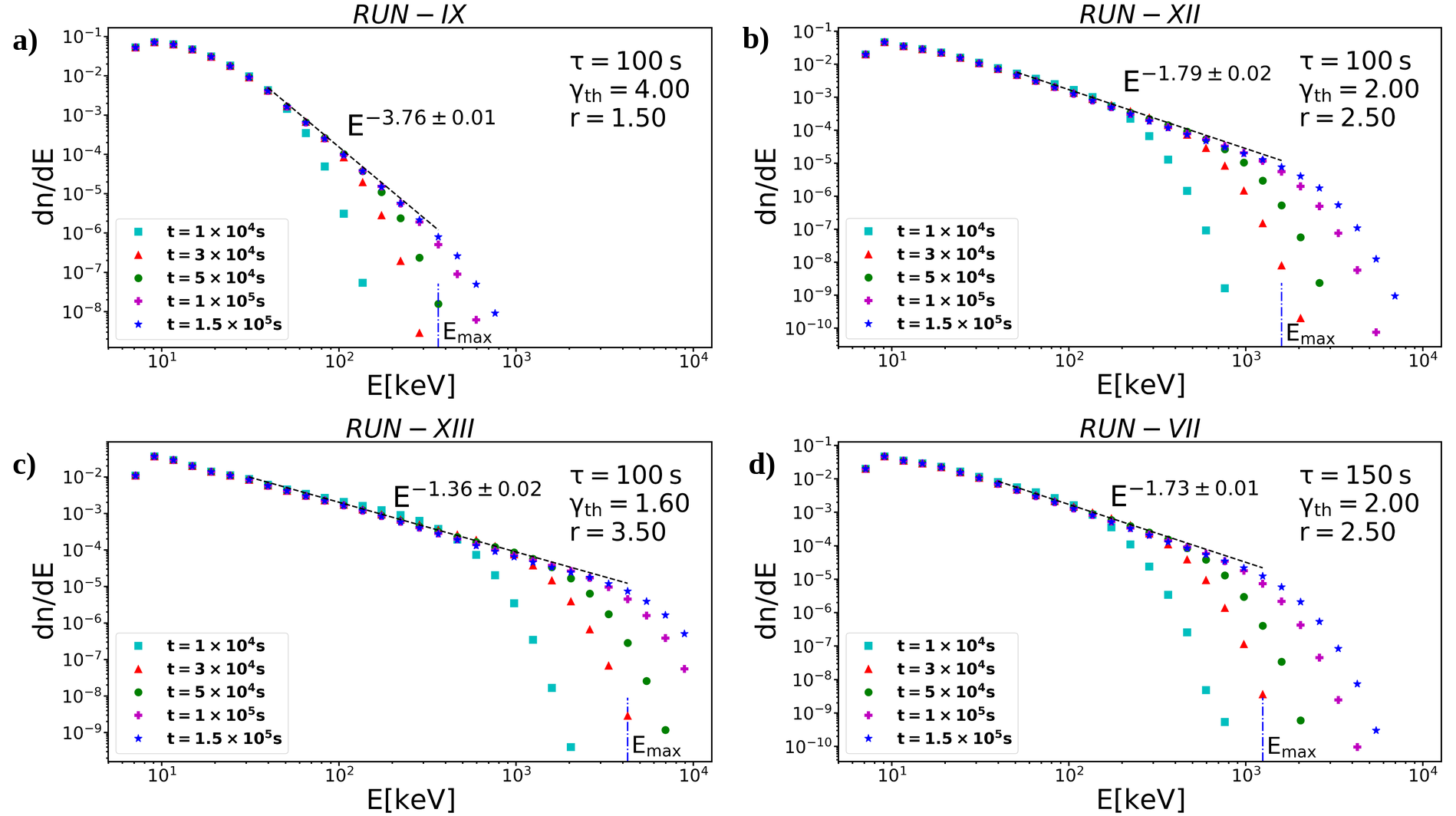}
  \caption{Energy spectra at different simulation times for DSA. In panels (a), (b), and (c) the compression ratio increases from $r=1.5$ to $r=3.5$, while keeping the scattering time $\tau=100$ s fixed. Panel (d) shows the same compression ratio as panel (b), but with $\tau = 150$ s. The dashed lines represent the spectra's best power-law fits of the form $E^{-\gamma}${, with the $\gamma$ values and the corresponding uncertainties shown in the panels.} 
  The theoretical spectral index $\gamma_{th}$ for each compression ratio, given by Eq. (\ref{spectral_index}), is reported {in the legend} for comparison. The different symbols correspond to different elapsed times, as indicated in the legend. }
  \label{spectra_diffusion}
\end{figure*}

As a starting point of the analysis, we evaluated the particle energy spectra at different times for both DSA and superdiffusive shock acceleration (SSA) by binning particles according to their energy and computing the corresponding particle distributions. The resulting spectra are expected to exhibit different slopes depending on whether diffusive or superdiffusive transport is considered \citep{Zimbardo13,Effenberger25,Aerdker25}.  In particular, the differential energy spectrum scales as
\be
   \frac{dn}{dE} = E^{-\gamma},
\ee
where $\gamma$ only depends on the shock-compression ratio, $r,$ for diffusive transport and on  $r$ and the exponent  $\alpha$ in the case of superdiffusion. Specifically \citep{Perri12a,Zimbardo13},
\be
\gamma =
  \begin{cases}
\displaystyle     \frac{2r + 1}{2(r-1)} \quad\quad\quad\quad \: \; \;{\rm DSA}, \\
\displaystyle     \frac{3}{r-1}\frac{2-\alpha}{3 - \alpha} + 1 \quad \quad {\rm SSA}.
  \end{cases}
\label{spectral_index}  
\ee
The spectra were computed by counting particles within 30 logarithmically spaced energy bins ranging from 6.25 keV up to 10 MeV.
To quantify the acceleration results, we define the `maximum energy', $E_{max}$, as the energy value up to which the particle spectrum follows a power-law distribution, starting at lower energies, and then bends down. Clearly, there are particles with energies larger than $E_{max}$, but their number decreases exponentially.\footnote{We actually made fits of the energy distributions obtained by the simulation with the \citet{Ellison85} distribution, which has an exponential cut-off, but the fits were not entirely satisfactory.} 
For each run, $E_{max}$ was evaluated visually from the spectrum and used to characterize the efficiency of particle acceleration under different shock configurations. In the next subsections, this quantity is compared with the lower energy threshold of the REPE detector ($\sim 2$ MeV) to assess whether accelerated protons are expected to be detected from the instrument or not.
Additionally, we evaluated the percentage of protons, $N_{particles}$, that overcame the energy threshold. This percentage was calculated as the ratio between the number of particles with energies equal to or higher than 2 MeV and the number of particles still present in the simulation box. While $E_{max}$ provides information on the maximum energy reached in the acceleration process, $N_{particles}$ quantifies the percentage of particles detectable by REPE; thus, it gives an indication of the intensity of the energetic particles.

\subsection{Normal diffusion}
\begin{figure*}[ht!]
  \centering
  \includegraphics[width=0.85\linewidth]{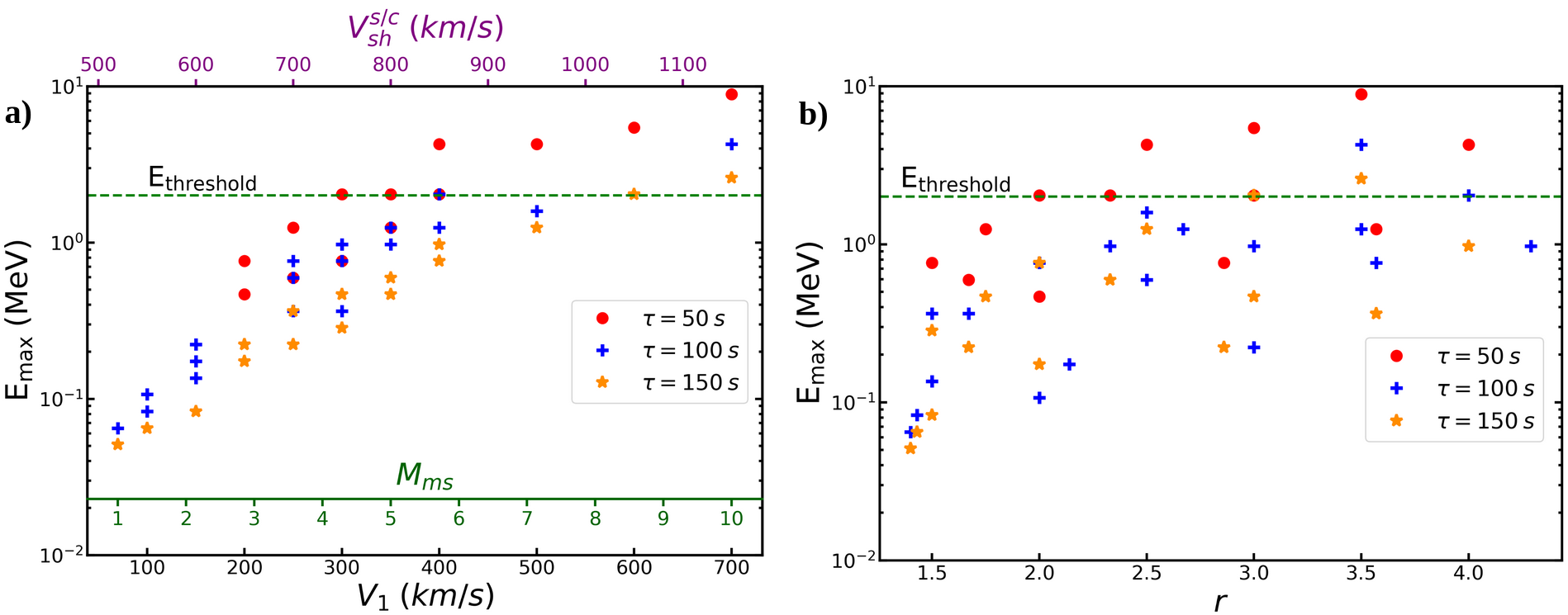}
  \caption{Maximum energy, $E_{max}$, as a function of the upstream velocity, $V_1$ (\textit{panel a}), and of the compression ratio, $r$ (\textit{panel b}), for different scattering times (indicated by the different symbols) and for different downstream velocities, $V_2$, the latter being given in the appendix. The horizontal dashed green line represents the lower energy threshold of the REPE instrument aboard HENON. In panel (a),  the horizontal solid green line indicates a reference magnetosonic Mach number; the upper horizontal axis also reports the shock speed as would be seen in the spacecraft frame. }
  \label{energy_max_diffusion}
\end{figure*}

\begin{figure}[ht!]
  \centering
  \includegraphics[width=0.87\linewidth]{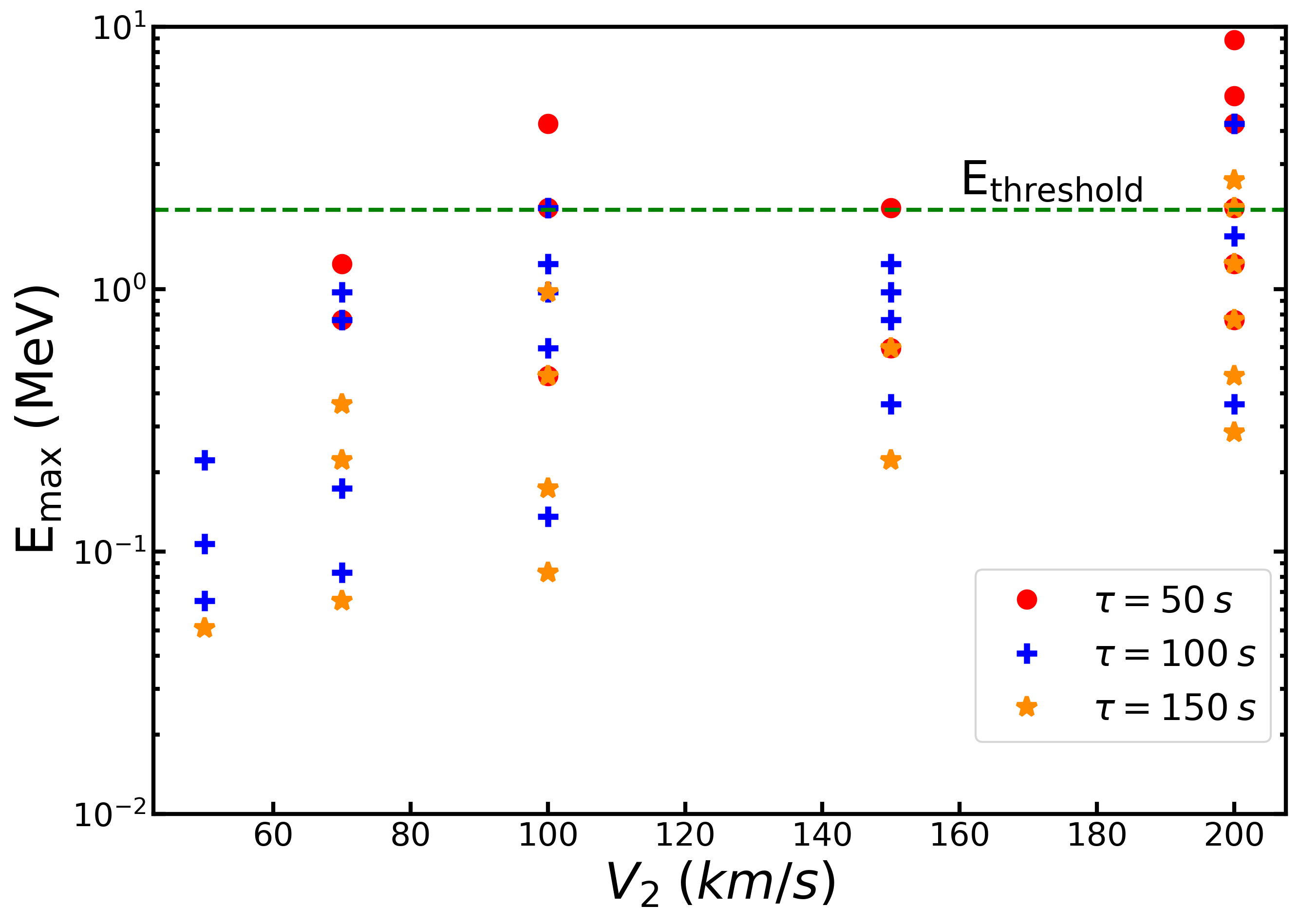}
  \caption{Maximum energy, $E_{max}$, as a function of the downstream velocity, $V_2$,  for different scattering times and different upstream velocities, $V_1$.  }
  \label{diff_energy_max_V2}
\end{figure}

The energy spectra for normal diffusion at different computational times are shown in Fig. \ref{spectra_diffusion}. As expected, a gradually more extended power-law spectrum is formed as time goes on.  
A power-law fit of the form $E^{-\gamma}${, with the corresponding uncertainty,} is over-plotted as a dashed line to compare the observed slope with the theoretical spectral index, $\gamma_{th}$. Because of the finite simulation time, even the power-law part of the spectrum is not a perfectly straight line in log--log axes, but it is slightly curved.
{We note that the particle power-law spectra are somewhat  harder than the theoretical predictions. This discrepancy can be associated with the choice of the power-law fitting range, which was selected by eye.}
The energy at which the spectrum at time of $t=1.5\times 10^5$ s deviates from the fitted power-law is identified as the maximum energy, $E_{max}$ (the values of $E_{max}$ for each run are reported in the tables in the appendix). 
Panels (a)--(c) show the energy spectra for shocks characterised by the same scattering time, $\tau = 100 \: s$, and different compression ratio values: $r = 1.5, 2.5,$ and $3.5 $, respectively. Panel (d) is included for comparison with panel (b): both spectra were evaluated for shocks with $r = 2.5$, but in panel (d) the scattering time is increased to $\tau = 150 \:s$. 
The title of the plots refers to the run number for each scattering time, and more details on the respective shock parameters are reported in the tables in Appendix A.

{A comparison of} the spectra {shows} that an increase in the compression ratio is associated with a flatter energy distribution, leading the spectral cut-off $E_{max}$ to shift towards higher energies.
This behaviour is consistent with the predictions. Indeed, a larger compression ratio corresponds to a steeper velocity jump across the shock, allowing the particles to gain more energy during each shock-crossing cycle; see Eq. (\ref{eq:momentum}). Consequently, the acceleration process becomes more efficient, leading to flatter spectra and larger values of $E_{max}$. 
Focusing on panel (d) instead, we observe that an increase in the scattering time, $\tau$, leads to a lower $E_{max}$ and a {slightly} less populated high-energy tail. Since particles are scattered less frequently, they undergo fewer shock-crossing cycles during the simulation, resulting in a lower overall acceleration efficiency.

The overall DSA simulation results are compared in the cumulative plots presented in the following. 
Figure \ref{energy_max_diffusion} shows the maximum cut-off energy, $E_{max}$, for different values of scattering times as a function of the upstream velocity and the compression ratio, in panels (a) and (b), respectively. It is clearly seen that $E_{max}$
steadily grows with $V_1$; $E_{max}$ also grows with the compression ratio, although a larger scatter in the data points is obtained. As expected, lower scattering times are related to higher final energies.  
The REPE lower energy threshold, indicated by the dashed horizontal green line, is overcome for $V_1 \ge 300$ km/s in the case of $\tau = 50 \:s$, whereas higher upstream velocities (stronger shocks) are needed for larger values of the scattering time. A reference magnetosonic Mach number, $M_{\rm ms}$, and the shock speed in the spacecraft frame, $V^{\rm s/c}_{\rm sh}$, are included in Fig. \ref{energy_max_diffusion}a, for qualitative comparison with measured shock parameters.  
Their values were estimated by assuming reference values of representative solar-wind conditions; 
in particular, a mean magnetosonic speed of $V_{\rm ms} = 70$ km/s and an upstream solar-wind speed (in the spacecraft frame) of $V_1^{\rm s/c} = 450$ km/s, similar to those reported in the statistical survey of \citet{Tang25}, were recorded. 

Panel (b) also exhibits an increasing trend with the compression ratio, even if it is less marked than panel (a). This confirms that stronger shocks lead to more efficient particle acceleration.

In contrast, we did not find a clear trend in Fig. \ref{diff_energy_max_V2}, which shows the $E_{max}$ as a function of the downstream velocity, $V_2$. This can be understood as the competing effects of the momentum gain, which decreases with the increase of $V_2$, as shown by Eq. (\ref{eq:momentum}), and of the cycle time given by Eq. (\ref{time_cycle}), which decreases with the increase of $V_2$. Therefore, $\Delta p/\Delta t$ varies only slowly with $V_2$.

\begin{figure*}[ht!]
  \centering
  \includegraphics[width=0.85\linewidth]{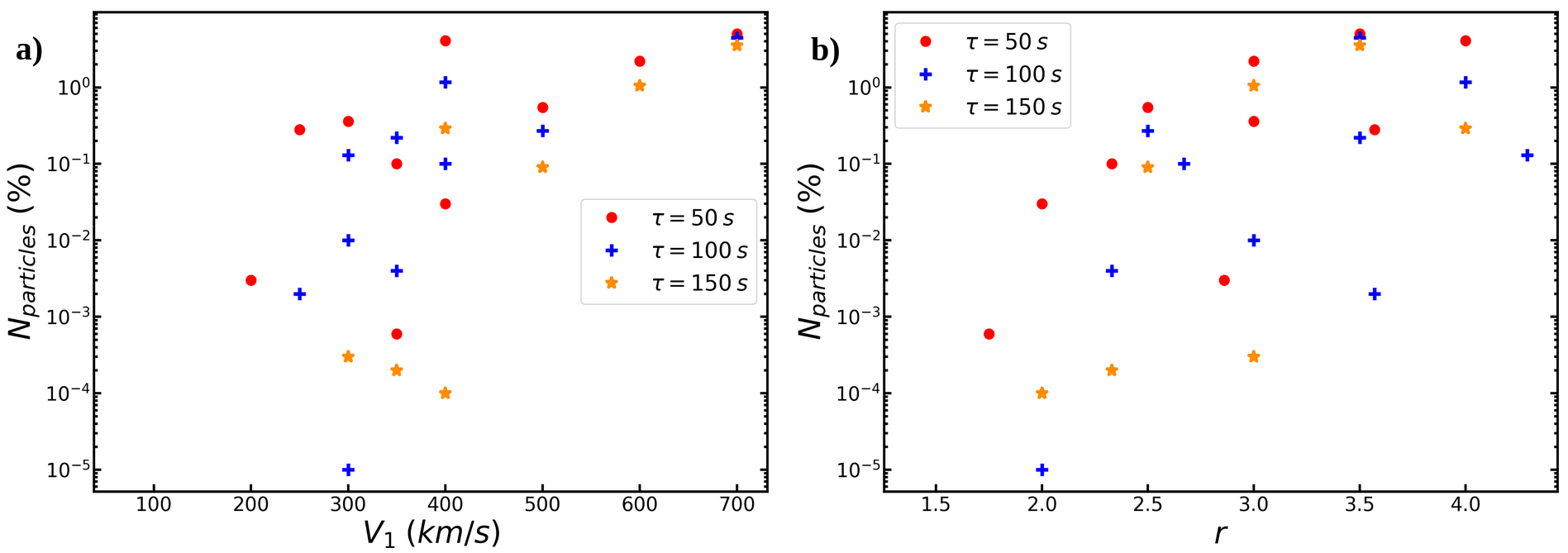}
  \caption{ Percentage of protons that overcome the 2 MeV threshold, $N_{particles}$, as a function of the upstream velocity, $V_1$ (\textit{panel a}), and of the compression ratio, $r$ (\textit{panel b}), for different scattering times.}
  \label{Nparticles_diffusion}
\end{figure*}

\begin{figure*}
  \centering
  \includegraphics[width=0.95\linewidth]{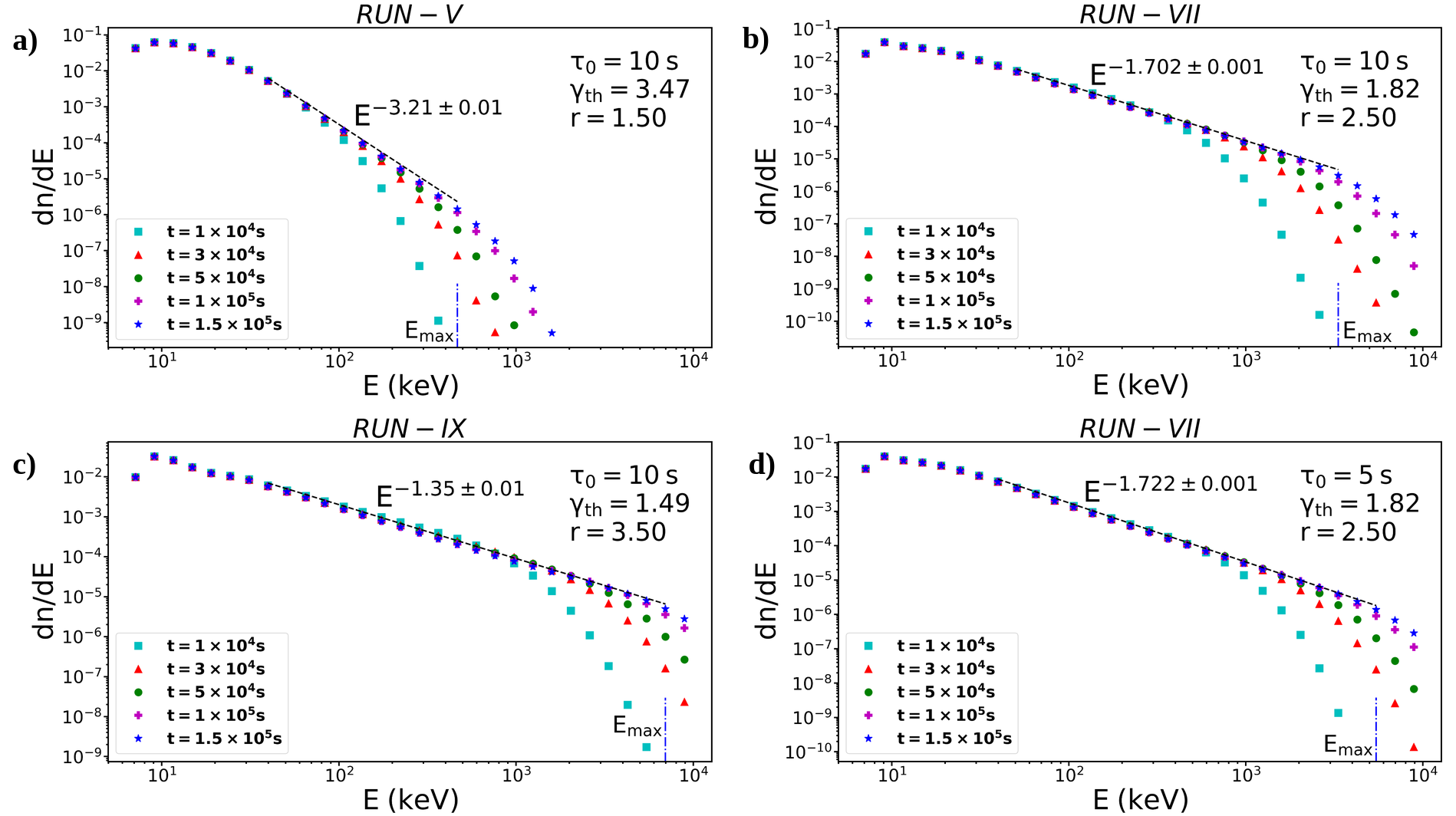}
  \caption{Energy spectra at different simulation times, indicated by the different symbols, for SSA. In panels (a), (b), and (c) the compression ratio increases from $r=1.5$ to $r=3.5$, while keeping $\tau_0=10$ s fixed. Panel (d) shows the same compression ratio as panel (b), but with $\tau_0 = 5$ s. The dashed lines represent the spectra's best power-law fits of the form $E^{-\gamma}${, with the $\gamma$ values and the corresponding uncertainties shown in the panels.}
  The theoretical spectral index, $\gamma_{th}$, is reported in the legend for comparison with that obtained {by the best power-law fit of the simulation results.} }
  \label{spectra_superdiffusion}
\end{figure*}
\begin{figure*}[ht!]
  \centering
  \includegraphics[width=0.85\linewidth]{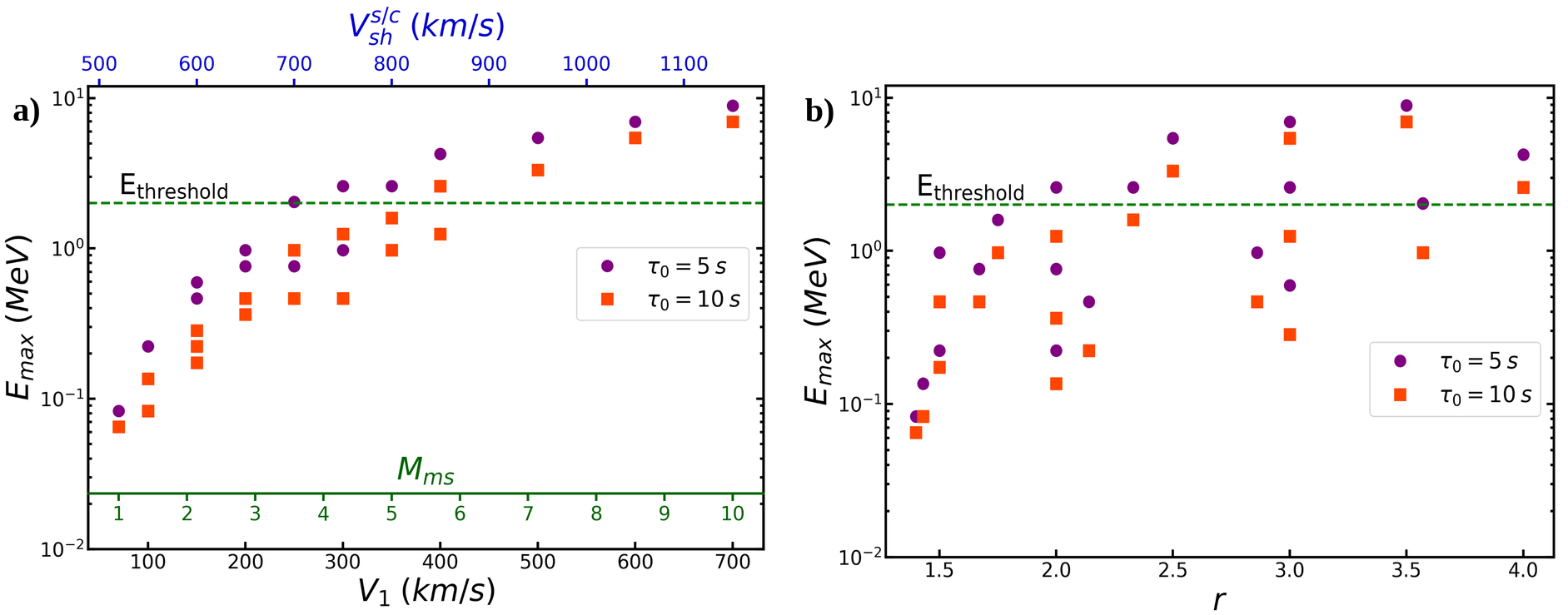}
  \caption{Maximum energy, $E_{max}$, as a function of the upstream velocity $V_1$ (\textit{panel a}) and of the compression ratio, $r$ (\textit{panel b}), for different scale times, $\tau_0$, as indicated in the legends. The downstream speed, $V_2,$ was also varied and is given in the tables in the appendix. The format is the same as Fig. \ref{energy_max_diffusion}. } 
  \label{energy_max_superdiffusion}
\end{figure*}
Finally, a comparison of the simulation results in terms of the percentage of protons whose energies reach or exceed the lower energy threshold of 2 MeV is reported in Fig. \ref{Nparticles_diffusion}. Panel (a) presents the results as a function of the upstream velocity, while panel (b) shows its dependence on the compression ratio, both for different scattering times and downstream velocities. Lower scattering times result in a larger percentage of particles exceeding the detection threshold, reflecting the higher acceleration efficiency. Moreover, the percentages of energetic particles increase with both the upstream velocity and the compression ratio. 
These trends are consistent with those observed for the spectral cut-off energy, confirming that faster and stronger shocks accelerate a larger fraction of particles to energies detectable by REPE. However, the scatter in the data points is large, i.e. up to four orders of magnitude for $V_1=300$--400 km/s and for compression ratios ($r)$ between two and three. This is reminiscent of the fact that the intensities of energetic storm particles vary over several orders of magnitude \citep[e.g.][]{Lee12}, arguably depending on several factors, so there is no single parameter allowing one to predict the intensity.

\subsection{Superdiffusion}
We also investigated particle acceleration across shocks in the case of superdiffusive transport. We repeated the numerical simulations using the same upstream and downstream velocities (i.e. same compression ratios) adopted in the case of DSA for a direct comparison between the two transport regimes.
The resulting superdiffusion energy spectra for $\alpha=1.3$ are shown in Fig. \ref{spectra_superdiffusion}. Panels (a), (b), and (c) compare the spectra with different compression ratios, $r = 1.5, 2.5,$ and $3.5$, respectively, with the same scale time of $\tau_0 = 10 \:s$. Panel (d) shows the energy spectrum with the same compression ratio as in panel (b), $r = 2.5 \:s$, but with a different scale time of $\tau_0 = 5 \:s$. 
As in the DSA case, an increase in the compression ratio produces progressively flatter spectra and higher maximum energies. However, for a given compression ratio, the superdiffusive spectra show energies somewhat higher than those of their DSA counterparts, as can be seen by contrasting Figs. \ref{spectra_diffusion} and \ref{spectra_superdiffusion}, indicating that superdiffusive transport enhances the acceleration efficiency. Also, superdiffusion generates harder spectral indices (see Eq. \ref{spectral_index} and \citealt{Prete26b}), a result that is particularly evident in Fig. \ref{spectra_superdiffusion}a. Furthermore, the power-law spectra are formed more quickly; for example, by looking at the turquoise squares, it can be seen that almost two decades of power-law spectrum are already formed at $\sim 10^4$ s, except for panel a). 
\begin{figure*}[ht!]
  \centering
  \includegraphics[width=0.85\linewidth]{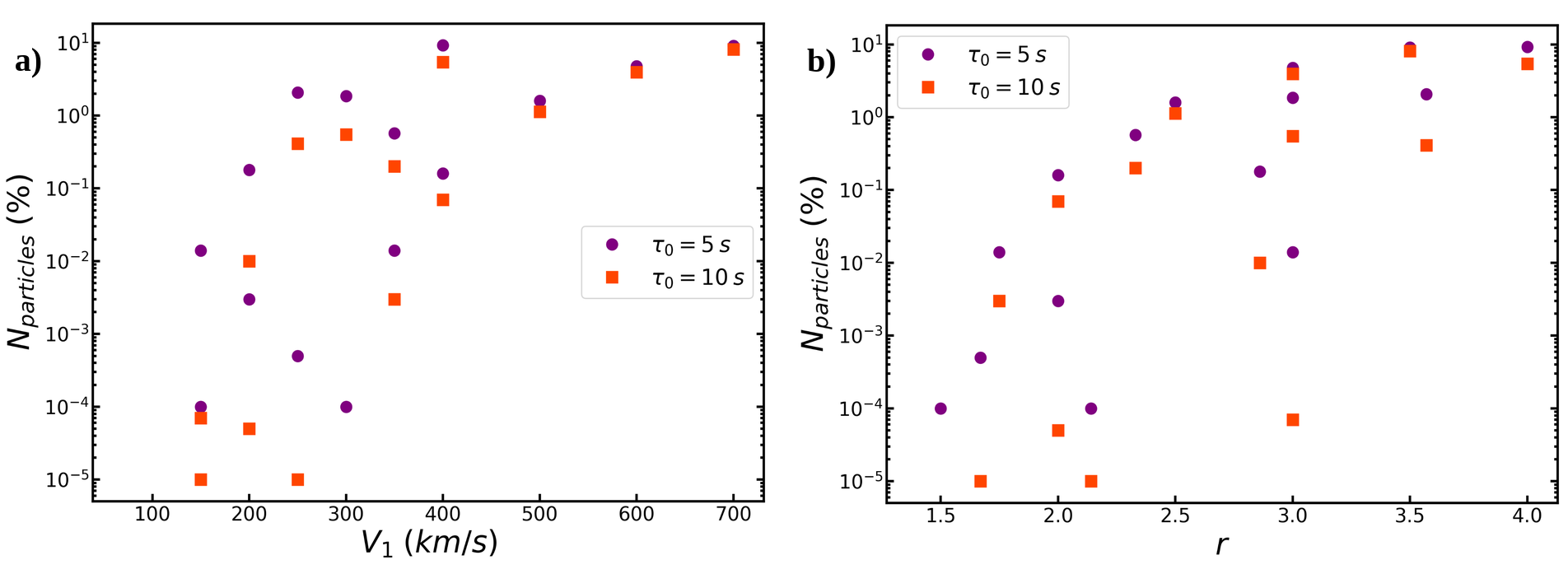}
  \caption{Percentage of protons that overcome the 2 MeV threshold, $N_{particles}$, as a function of the upstream velocity, $V_1$ (\textit{panel a}), and the compression ratio, $r$ (\textit{panel b}), for different {values of the minimum scattering time,} $\tau_0$.}
  \label{Nparticles_superdiffusion}
\end{figure*}
Next, we evaluated the maximum energy, $E_{max}$, as a function of the upstream velocity and the compression ratio, for $\tau_0 = 5$ s and 10 s, shown in Fig. \ref{energy_max_superdiffusion}. Similarly to the DSA case, we reported reference values of the magnetosonic Mach number and the shock speed in the spacecraft frame. The maximum energy follows a clear increasing trend with the upstream velocity, and the energy values are greater for lower scattering times. {The rate of increase of $E_{max}$ as a function of the upstream speed is larger than in the case of DSA.}
Panel (b\index{b}) also shows an increasing trend of $E_{max}$ with the compression ratio, but, for a fixed value of $r$, the maximum energy is higher for larger upstream velocities (see the tables in Appendix A). As for normal diffusion, the energy values are more scattered when plotted as a function of $r$.

\begin{figure*}[ht!]
  \centering
  \includegraphics[width=0.95\linewidth]{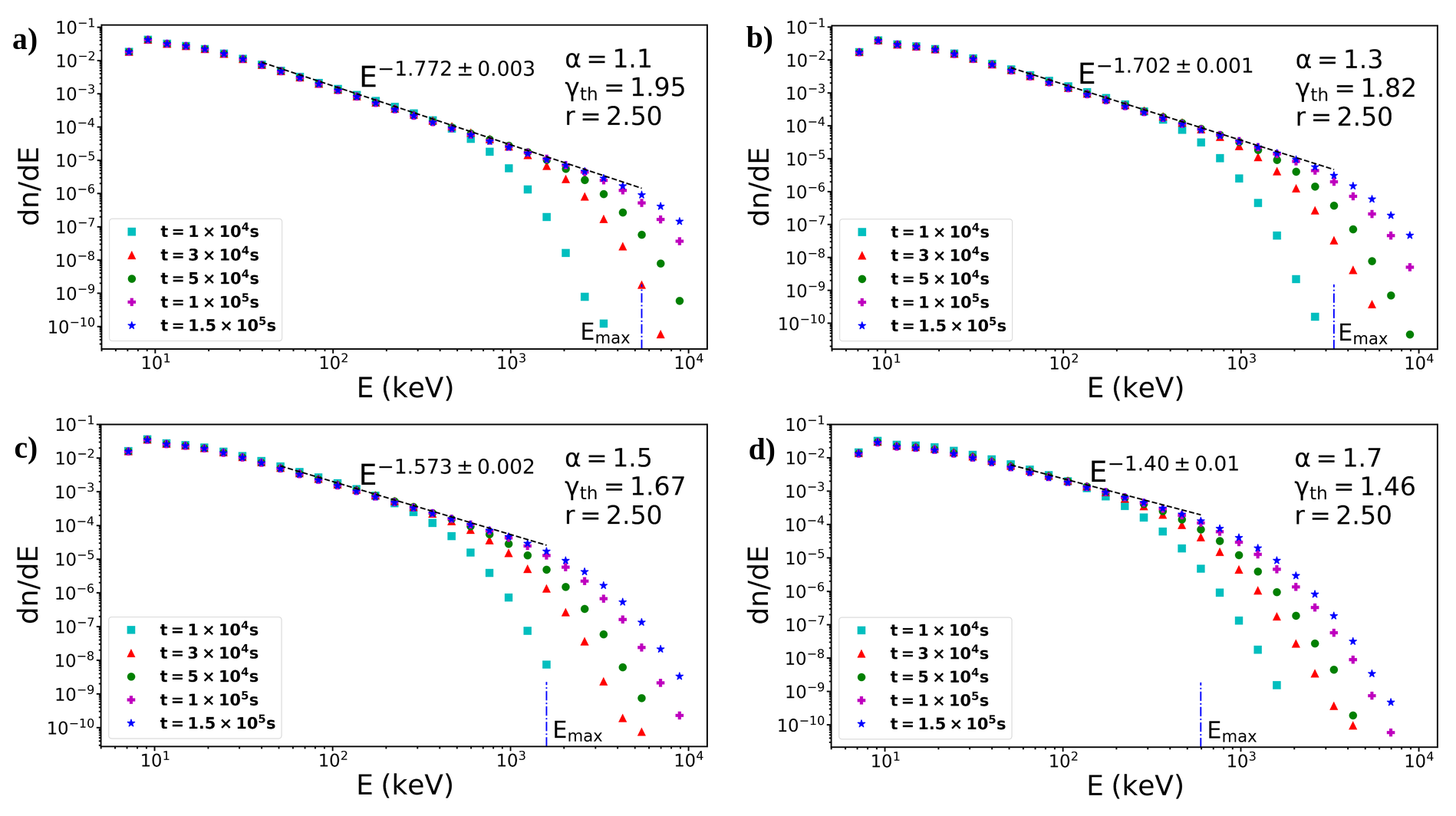}
  \caption{Energy spectra for the superdiffusive cases with different $\alpha$ values, as indicated in the legend. All runs have the same compression ratio, $r=2.5,$ and the same scale time, $\tau_0= 10$ s. The format is the same as Fig. 2.}
  \label{spectra_mu}
\end{figure*}
 
The percentage of protons that exceed the lower energy threshold of REPE is shown in Fig. \ref{Nparticles_superdiffusion} for two different scale times, $\tau_0$. 
Both panels show an increase in the percentage with the upstream velocity and the compression ratio, similar to the normal diffusive case. However, it can be seen that in the superdiffusive case, even some low-speed shocks, with $V_1=150$ km/s, are able to accelerate protons above 2 MeV. 
\begin{figure*}[ht!]
  \centering
  \includegraphics[width=0.85\linewidth]{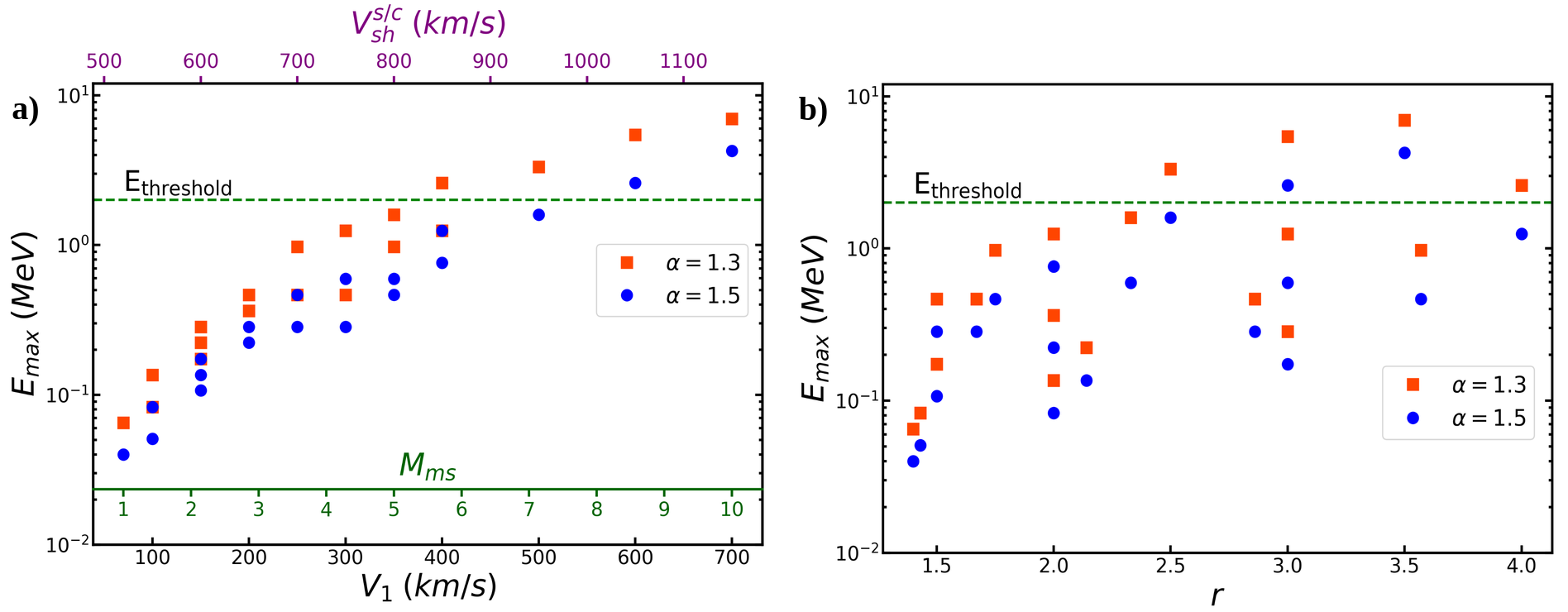}
  \caption{ Maximum energy, $E_{max}$, as a function of the upstream velocity, $V_1$ (\textit{panel a}), and the compression ratio, r (\textit{panel b}), for different $\alpha$ values. The format is the same
as Fig. \ref{energy_max_diffusion}.}
  \label{mu_energy}
\end{figure*}
\begin{figure*}[ht!]
  \centering
  \includegraphics[width=0.85\linewidth]{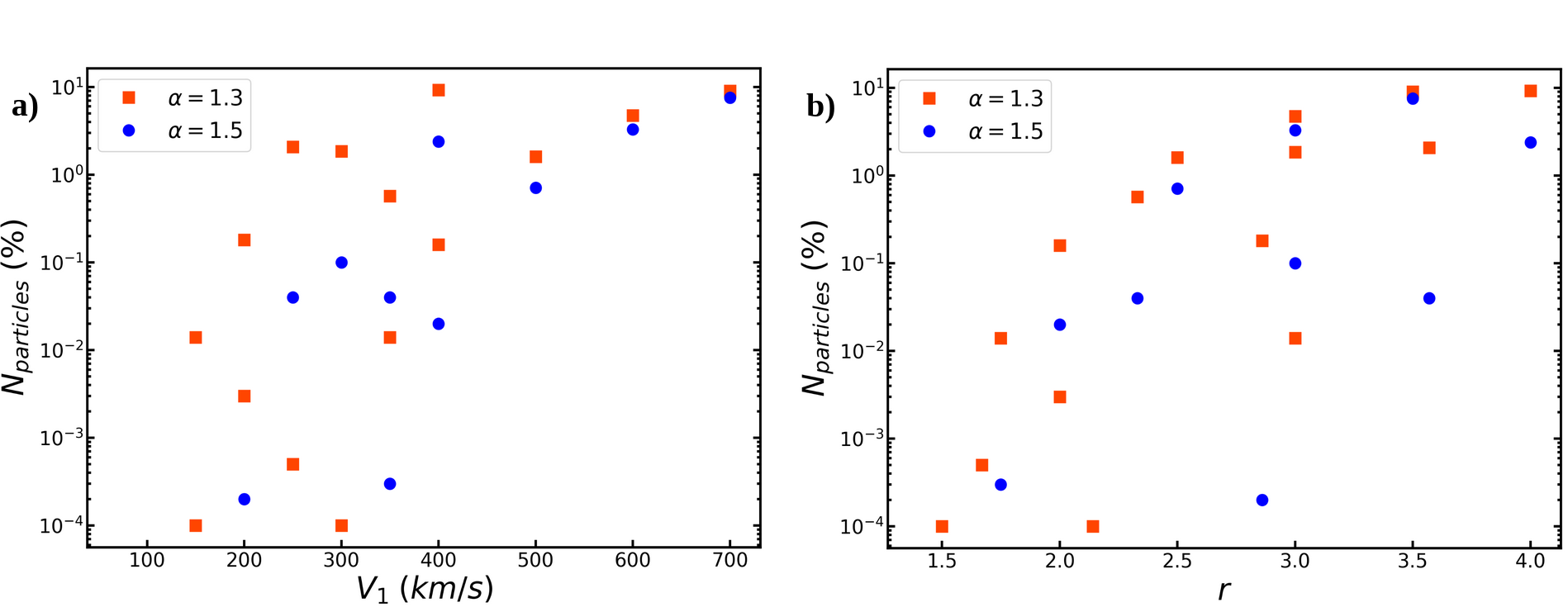}
  \caption{ Percentage of protons that overcome the 2 MeV threshold, $N_{particles}$, as a function of the upstream velocity, $V_1$ (\textit{\textit{panel a}}), and of the compression ratio, $r$ (\textit{panel b}), for different $\alpha$ values, as indicated in the legend, and for $\tau_0=10$ s.}
  \label{mu_particles}
\end{figure*}

All the superdiffusive runs discussed so far were performed with a superdiffusive exponent $\alpha = 1.3$. However, the value of $\alpha$ is not fixed a priori, but it can vary in the $1<\alpha<2$ range; the study of the upstream profile of several shock crossings in the heliosphere indicates that in most cases $1.3<\alpha<1.6$ \citep{Perri15,Perri15b,Zimbardo18}. 
{To investigate the effect of the anomalous transport on particle acceleration, we performed a further set of simulations varying the superdiffusion exponent  $\alpha$, while keeping $\tau_0 = 10$ s and $r=2.5$. The particle energy spectra for $\alpha= 1.1, 1.3, 1.5,$ and 1.7 are shown in Fig. 9. As can be seen from this figure, an increase in the superdiffusive exponent $\alpha$ leads to harder spectra, in agreement with the steady-state SSA theory and Eq. (\ref{spectral_index}). However, for a finite shock lifetime the maximum energy, $E_{max}$, decreases significantly, owing to the increase of $T_{\rm cycle}$ (shown in Eq. (\ref{time_cycle_super})) with the $\alpha$ parameter, as discussed in Sect. 2. Yet, shorter acceleration times can still be obtained for smaller values of $\tau_0$ (see also the discussions in \citealt{Perri15} and \citealt{Prete26b}). These results emphasize that the maximum reachable energy depends on the total available time for acceleration (e.g. the shock lifetime) and the acceleration cycle time, which depends sensitively on the transport properties. }

{In order to compare with the energy threshold of the HENON instrument,} we performed another set of runs with $\alpha = 1.5$ and $\tau_0 = 10$ s, while keeping all the other parameters fixed. Figures \ref{mu_energy} and \ref{mu_particles} compare the maximum energy, $E_{max}$, and the percentage of particles, $N_{particles}$, as a function of $V_1$ and $r$. The results confirm that an increase in the exponent of superdiffusion leads to a lower particle acceleration efficiency. Indeed, the results with $\alpha = 1.5$ have lower values of $E_{max}$ with respect to $\alpha = 1.3$, and a smaller fraction of particles above the detection threshold.
This suggests that faster superdiffusive transport, which is related to larger values of $\alpha$, allows particles to spread away from the shock more rapidly. As a consequence, particles cross the shock less frequently, resulting in lower maximum energies despite the enhanced spatial transport. 
We also notice from Fig. \ref{mu_energy} that when going from $\alpha=1.3$ to $\alpha=1.5$ the maximum energy decreases by a factor of two or less. Conversely, the influence of changing $\alpha$ is stronger on the number of protons that reach the 2 MeV threshold; as shown by  Fig. \ref{mu_particles}, the percentages can decrease by an order of magnitude and sometimes more. Once again, the intensity of energetic storm particles can strongly depend on the transport parameters.

\section{Summary and discussion}

In this work, we used the test particle simulation by \citet{Prete26b} to study proton acceleration at interplanetary shocks in order to shed light on both the fundamental issue of shock acceleration in the heliosphere and the possibility for the HENON spacecraft to observe energetic particles with its REPE instrument, which has a minimum energy threshold of 2 MeV. We made several runs with simulation parameters (bulk velocities, scattering times, compression ratios, and shock lifetimes) modelled according to IP shock observations. Both normal diffusion and superdiffusion were considered for the particle-scattering process. The latter case requires generating a power-law distribution of scattering times, with the power-law exponent related to the superdiffusion exponent by $4-\alpha$, as shown in Eq. (\ref{eq:levyWalk}).

The same {exponential proton population shown in Fig. \ref{init_E0} was injected within the simulation box for all runs}. In particular, we did not correlate the features of the injected distribution with the shock parameters for each run. In other words, we did not address the injection problem of DSA. Of course, from a physical point of view, it is reasonable to think that higher Mach number shocks are able to create more abundant suprathermal ions. This happens, for example, via the process of thermal ion reflection; i.e. the number of reflected ions increases with the Mach number and the plasma, $\beta$ \citep[e.g.][]{Leroy82,Edmiston84,Laming22}, and these ions can undergo pre-acceleration at the shock front. However, such a study would require a self-consistent particle simulation; we plan to do this in a future work. 

The simulation results show that IP shocks are able to generate a power-law energy distribution rather quickly (in about 6 h for normal diffusion and in about 3 h for superdiffusion), and that the obtained energy spectral indices are in good agreement with the theoretical expectations. For typical parameters, superdiffusion turns out to generate somewhat faster acceleration. 
For practical purposes, we define the maximum energy {as the energy at which the high energy cut-off in the particle spectrum arises.}
For all the {performed runs and for all the types of transport considered,} we find that $E_{max}$ grows with the shock upstream velocity, $V_1,$ and the compression ratio, $r$, with a clearer and more ordered dependence on $V_1$. {The latter result is in agreement with the fact that the momentum gain at each shock crossing increases with $V_1$ (as shown by Eq. \ref{eq:momentum}) and the cycle time around the shock decreases with $V_1$ (as shown by Eqs. \ref{time_cycle} and \ref{time_cycle_super}).} 
In addition, reaching 2 MeV also depends on the scattering time, $\tau,$ for normal diffusion, and on the scale time, $\tau_0$, and exponent, $\alpha$, for superdiffusion.

The percentage of protons above 2 MeV, which reflects the intensity of energetic storm particles, varies substantially --- by up to five orders of magnitude for a given upstream velocity, $V_1$ --- as has been found in many heliospheric shock observations. This large variability depends only on the simulation parameters, the proton injection being the same for all runs. On the other hand, we can imagine that in the IP space the intensity of energetic particles also depends on the presence of a seed population of suprathermal particles, whose abundance may depend on both the shock strength and the history of the heliosphere. Therefore, other reasons for variability can exist besides the parameter variations in our simulations, so the intensity of ESP remains difficult to constrain. 

In summary, for both diffusion and superdiffusion, as a rough estimate we can say that the HENON energy threshold of 2 MeV is reached when the shock speed in the spacecraft frame is above 800 km/s. Conversely, when little or no acceleration above 100 keV is found, this is probably due to the low bulk speed, $V_1$, in the shock frame, as clearly suggested by Figs. \ref{energy_max_diffusion},  \ref{energy_max_superdiffusion}, and \ref{mu_energy}. 

We can compare our numerical results with some studies of ESPs in the heliosphere. Using STEREO data, \citet{Dresing16} found that about 27\% of shocks exhibit an increase of 100 keV ions, and only 5\% of shocks exhibit an increase in the 2 MeV ions. A similar trend, although with different percentages, is found for the IP shocks observed by the Solar Orbiter \citep[][Fig. 9]{Trotta25b}. Based on our results, the events with no 100 keV ion enhancements can be interpreted as $V_1$ shock speeds less than 100 km/s in the shock frame (i.e. smaller than about 550 km/s in the spacecraft frame, remembering that $V_{\rm sh}^{\rm s/c} = V_1^{\rm s/c} + |V_1|$), corresponding to magnetosonic Mach numbers of $M_{\rm ms}< 1.5$.  
These inferences are consistent with the statistical study by \citet{Trotta25b}, which found that the shock speeds in the spacecraft frame are clustered around 500 km/s, so moderate speeds of $V_1$ are the most frequent. Also, \citet{Trotta25b} found that $M_{\rm ms}$ has a frequency distribution clustered around small values such as 1.5 and 2. Furthermore, the small percentage of shocks (5\%) with an increase in the 2 MeV flux, found by \citet{Dresing16}, can be explained by the fact that relatively large velocities of $V_1$ are needed to reach this energy, with $V_{\rm sh}^{\rm s/c}>800$ km/s, but such velocities are not very frequent, as shown by \citet{Trotta25b}, Figs. 2 and 4. 
It is interesting to note that this `threshold velocity' of 800 km/s was also found by \citet{Kartavykh25} as the shock speed above which so-called plateau profiles for the intensity of energetic protons are mostly found, possibly in association with proton mean free paths of less than 0.015 au (as discussed in Sect. 2, this mean free path corresponds to scattering times of 150 s). The finding of \citet{Kartavykh25} concerning plateau profiles indicates that fast shocks induce normal diffusion rather than superdiffusion (which may correspond to the spike profiles); this can be tentatively interpreted as the fact that fast shocks 
accelerate particles to larger energies and also accelerate more particles than slower shocks: more energetic particles can excite magnetic turbulence close to the shock, which in turn favours quick particle scattering and therefore normal transport \citep{Perri23}. 

Regarding space-weather predictions, when energetic protons are observed by the REPE instrument onboard HENON, based on our simulation results we can infer that the shock speed of $V_{\rm sh}^{\rm s/c}$ is of the order of or above 800 km/s. When a fast CME propagates away from the Sun, the shock speed can be considered as an estimate of the driving CME speed, $V$, which in turn enters the $VB_z$ parameter, with $B_z$ being the `southward' component of the interplanetary magnetic field. This parameter corresponds to the motional electric field of the solar wind and is considered to be one of the main drivers of strong geomagnetic storms \citep{Gonzalez89,Wang03,Prete26a}. Therefore, the detection of energetic protons by REPE in association with an IP shock can be used as a proxy for the CME speed and can also give a precious indication when the plasma speed is not available from the plasma instrument.
This information will mostly be probabilistic because many parameters can influence the particle maximum energy and intensity, as discussed above. Still, the detection of protons with $E>2$ MeV by HENON can be useful for issuing a pre-alert, taking advantage of the fact that a CME-driven shock precedes the CME by several  hours. 

\begin{acknowledgements}
        This work is the outcome of the preparatory phases for the launch of the HENON mission by the Italian Space Agency (ASI) and was partially funded through the Argotec contracts, numbers ARG-IT-CON-P-HEN-220002 and ARG-IT-CON-P-HEN-250003. HENON is part of the ASI programme Alcor and is being developed under the European Space Agency's General Support Technology Programme (ESA-GSTP) through the support of the national delegations of Italy (ASI), the UK, Finland, and the Czech Republic.
\end{acknowledgements}

\bibliographystyle{bibtex/aa} 
\bibliography{henon2} 

\clearpage

\appendix
\section{Simulation parameters}
\label{appendix}
\nolinenumbers

The simulation parameters adopted for both the DSA and SSA models are summarised in the tables below. Each table refers to a set of simulations performed with a fixed scattering time, $\tau$ for DSA and $\tau_0$ for SSA.
In the DSA case, we performed simulations with $\tau = 50, 100, 150 \:s$, while the SSA model was investigated with $\tau_0 = 5, 10 \:s$. 
For each simulation, the tables show the Run number, the upstream and downstream velocities {in the shock frame}, the compression ratio $r$, the theoretical spectral index $\gamma_{{\rm th}}$, the maximum energy $E_{max}$, the percentage of particles with energies above the lower energy threshold of REPE, $N_{particles}$.
The quantity $E_{max}$ is evaluated from the energy spectra at the last simulation time analysed ($ t = 1.5 \times 10^5 \: s$), while the particle percentage, $N_{particles}$, is calculated with respect to the total number of particles remaining in the simulation domain at that time.

\subsection{Normal diffusion}

\renewcommand{\arraystretch}{1.15}
\setlength{\tabcolsep}{6pt}

\begin{table}[ht!]
\centering
\caption{Simulation parameters and resulting proton detections for the DSA model with $\tau = 50 \:s$.}
\label{RUNS}

\begin{tabular}{ccccccc}
\hline\hline  
Run & $V_1$ & $V_2$ & $r$ & $\gamma_{\rm th}$ & $E_{max}$ & $N_{particles}$ \\
& [km/s] & [km/s] & & & [MeV] & [\%] \\
\hline

I     & 200 &  70 & 2.86 & 1.81 & 0.762 & 0.003 \\
II    & 200 & 100 & 2.00 & 2.50 & 0.466 &  \textemdash \\
III   & 300 & 100 & 3.00 & 1.75 & 2.037 & 0.36 \\
IV    & 400 & 100 & 4.00 & 1.50 & 4.261 & 4.08 \\
V     & 300 & 200 & 1.50 & 4.00 & 0.762 &  \textemdash \\
VI    & 400 & 200 & 2.00 & 2.50 & 2.037 & 0.03 \\
VII   & 500 & 200 & 2.50 & 2.00 & 4.261 & 0.55 \\
VIII  & 600 & 200 & 3.00 & 1.75 & 5.448 & 2.20 \\
IX    & 700 & 200 & 3.50 & 1.60 & 8.910 & 5.00 \\
X     & 250 &  70 & 3.57 & 1.58 & 1.246 & 0.28 \\
XI    & 250 & 150 & 1.67 & 3.25 & 0.596 & \textemdash \\
XII   & 350 & 150 & 2.33 & 2.12 & 2.037 & 0.10 \\
XIII  & 350 & 200 & 1.75 & 3.00 & 1.246 & 0.0006 \\

\hline  
\end{tabular}

\end{table}

\renewcommand{\arraystretch}{1.15}
\setlength{\tabcolsep}{6pt}

\begin{table}[ht!]
\centering
\caption{Simulation parameters and resulting proton detections for the DSA model with $\tau = 100 \: s$.}
\label{RUNS_t100}

\begin{tabular}{ccccccc}
\hline\hline  
Run & $V_1$ & $V_2$ & $r$ & $\gamma_{\rm th}$ & $E_{max}$ & $N_{particles}$ \\
& [km/s] & [km/s] & & & [MeV] & [\%] \\
\hline

I      & 250 &  70 & 3.57 & 1.58 & 0.762 & 0.002 \\
II     & 300 &  70 & 4.29 & 1.46 & 0.974 & 0.13 \\
III    & 250 & 100 & 2.50 & 2.00 & 0.596 & \textemdash \\
IV     & 250 & 150 & 1.67 & 3.25 & 0.364 &\textemdash \\
V      & 300 & 100 & 3.00 & 1.75 & 0.974 & 0.01 \\
VI     & 300 & 150 & 2.00 & 2.50 & 0.762 & 0.00001 \\
VII    & 350 & 100 & 3.50 & 1.60 & 1.246 & 0.22 \\
VIII   & 350 & 150 & 2.33 & 2.12 & 0.974 & 0.004 \\
IX     & 300 & 200 & 1.50 & 4.00 & 0.364 & \textemdash \\
X      & 400 & 100 & 4.00 & 1.50 & 2.037 & 1.17 \\
XI     & 400 & 150 & 2.67 & 1.90 & 1.246 & 0.10 \\
XII    & 500 & 200 & 2.50 & 2.00 & 1.593 & 0.27 \\
XIII   & 700 & 200 & 3.50 & 1.60 & 4.261 & 4.45 \\
XIV    & 150 &  50 & 3.00 & 1.75 & 0.223 & \textemdash \\
XV     & 150 &  70 & 2.14 & 2.31 & 0.174 & \textemdash \\
XVI    & 150 & 100 & 1.50 & 4.00 & 0.136 & \textemdash \\
XVII   & 100 &  70 & 1.43 & 4.50 & 0.083 & \textemdash \\
XVIII  & 100 &  50 & 2.00 & 2.50 & 0.107 & \textemdash \\
XIX    &  70 &  50 & 1.40 & 4.75 & 0.065 & \textemdash \\

\hline
\end{tabular}

\end{table}

\renewcommand{\arraystretch}{1.15}
\setlength{\tabcolsep}{6pt}

\begin{table}[ht!]
\centering
\caption{Simulation parameters and resulting proton detections for the DSA model with $\tau = 150 \: s$.}
\label{RUNS_t150}

\begin{tabular}{ccccccc}
\hline\hline  
Run & $V_1$ & $V_2$ & $r$ & $\gamma_{\rm th}$ & $E_{max}$ & $N_{particles}$ \\
& [km/s] & [km/s] & & & [MeV] & [\%] \\
\hline

I     & 200 &  70 & 2.86 & 1.81 & 0.223 & \textemdash \\
II    & 200 & 100 & 2.00 & 2.50 & 0.174 & \textemdash \\
III   & 300 & 100 & 3.00 & 1.75 & 0.466 & 0.0003 \\
IV    & 400 & 100 & 4.00 & 1.50 & 0.974 & 0.29 \\
V     & 300 & 200 & 1.50 & 4.00 & 0.285 & \textemdash \\
VI    & 400 & 200 & 2.00 & 2.50 & 0.762 & 0.0001 \\
VII   & 500 & 200 & 2.50 & 2.00 & 1.246 & 0.09 \\
VIII  & 600 & 200 & 3.00 & 1.75 & 2.037 & 1.05 \\
IX    & 700 & 200 & 3.50 & 1.60 & 2.605 & 3.53 \\
X     & 250 &  70 & 3.57 & 1.58 & 0.364 & \textemdash \\
XI    & 250 & 150 & 1.67 & 3.25 & 0.223 & \textemdash \\
XII   & 350 & 150 & 2.33 & 2.12 & 0.596 & 0.0002 \\
XIII  & 350 & 200 & 1.75 & 3.00 & 0.466 & \textemdash \\
XIV   & 150 & 100 & 1.50 & 4.00 & 0.083 & \textemdash \\
XV    & 100 &  70 & 1.43 & 4.50 & 0.065 & \textemdash \\
XVI   &  70 &  50 & 1.40 & 4.75 & 0.051 & \textemdash \\

\hline  
\end{tabular}

\end{table}

\subsection{Superdiffusion}

\renewcommand{\arraystretch}{1.15}
\setlength{\tabcolsep}{6pt}

\begin{table}[ht!]
\centering
\caption{Simulation parameters and resulting proton detections for the SSA model with $\tau_0 = 5 \:s$.}
\label{RUNS_5t0supdiff}

\begin{tabular}{ccccccc}
\hline\hline
Run & $V_1$ & $V_2$ & $r$ & $\gamma_{\rm th}$ & $E_{max}$ & $N_{particles}$ \\
& [km/s] & [km/s] & & & [MeV] & [\%] \\
\hline

I      & 200 &  70 & 2.86 & 1.67 & 0.974 & 0.18 \\
II     & 200 & 100 & 2.00 & 2.24 & 0.762 & 0.003 \\
III    & 300 & 100 & 3.00 & 1.62 & 2.605 & 1.85 \\
IV     & 400 & 100 & 4.00 & 1.41 & 4.260 & 9.19 \\
V      & 300 & 200 & 1.50 & 3.47 & 0.974 & 0.0001 \\
VI     & 400 & 200 & 2.00 & 2.24 & 2.605 & 0.16 \\
VII    & 500 & 200 & 2.50 & 1.82 & 5.448 & 1.60 \\
VIII   & 600 & 200 & 3.00 & 1.62 & 6.967 & 4.74 \\
IX     & 700 & 200 & 3.50 & 1.49 & 8.910 & 8.87 \\
X      & 250 &  70 & 3.57 & 1.48 & 2.037 & 2.07 \\
XI     & 250 & 150 & 1.67 & 2.85 & 0.762 & 0.0005 \\
XII    & 350 & 150 & 2.33 & 1.93 & 2.605 & 0.57 \\
XIII   & 350 & 200 & 1.75 & 2.65 & 1.593 & 0.014 \\
XIV    & 150 &  50 & 3.00 & 1.62 & 0.596 & 0.014 \\
XV     & 150 &  70 & 2.14 & 2.08 & 0.466 & 0.0001 \\
XVI    & 150 & 100 & 1.50 & 3.47 & 0.223 & \textemdash \\
XVII   & 100 &  70 & 1.43 & 3.88 & 0.136 & \textemdash \\
XVIII  & 100 &  50 & 2.00 & 2.23 & 0.223 & \textemdash \\
XIX    &  70 &  50 & 1.40 & 4.09 & 0.083 & \textemdash \\

\hline
\end{tabular}

\end{table}

\renewcommand{\arraystretch}{1.15}
\setlength{\tabcolsep}{6pt}

\begin{table}[ht!]
\centering
\caption{Simulation parameters and resulting proton detections for the SSA model with $\tau_0 = 10 \: s$.}
\label{RUNS_10t0supdiff}

\begin{tabular}{cccccccc}
\hline\hline
Run & $V_1$ & $V_2$ & $r$ & $\gamma_{\rm th}$ & $E_{max}$ & $N_{particles}$ \\
& [km/s] & [km/s] & & & [MeV] & [\%] \\
\hline
I      & 200 &  70 & 2.86 & 1.67 & 0.466 & 0.01 \\
II     & 200 & 100 & 2.00 & 2.24 & 0.364 & 0.00005 \\
III    & 300 & 100 & 3.00 & 1.62 & 1.246 & 0.55 \\
IV     & 400 & 100 & 4.00 & 1.41 & 2.605 & 5.44 \\
V      & 300 & 200 & 1.50 & 3.47 & 0.466 & \textemdash \\
VI     & 400 & 200 & 2.00 & 2.24 & 1.246 & 0.07 \\
VII    & 500 & 200 & 2.50 & 1.82 & 3.332 & 1.13 \\
VIII   & 600 & 200 & 3.00 & 1.62 & 5.448 & 3.96 \\
IX     & 700 & 200 & 3.50 & 1.49 & 6.967 & 8.04 \\
X      & 250 &  70 & 3.57 & 1.48 & 0.974 & 0.41 \\
XI     & 250 & 150 & 1.67 & 2.85 & 0.466 & 0.00001 \\
XII    & 350 & 150 & 2.33 & 1.93 & 1.593 & 0.20 \\
XIII   & 350 & 200 & 1.75 & 2.65 & 0.974 & 0.003 \\
XIV    & 150 &  50 & 3.00 & 1.62 & 0.285 & 0.00007 \\
XV     & 150 &  70 & 2.14 & 2.08 & 0.223 & 0.00001 \\
XVI    & 150 & 100 & 1.50 & 3.47 & 0.174 & \textemdash \\
XVII   & 100 &  70 & 1.43 & 3.88 & 0.083 & \textemdash \\
XVIII  & 100 &  50 & 2.00 & 2.24 & 0.136 & \textemdash \\
XIX    &  70 &  50 & 1.40 & 4.09 & 0.065 & \textemdash \\
\hline
\end{tabular}

\end{table}


\end{document}